\documentclass[11pt]{article}
\usepackage{fullpage}

\usepackage{amsmath,amssymb,braket,color,dcolumn}
\usepackage{qcircuit}

\usepackage{hyperref}

\title{Multipartite High-dimensional Quantum State Engineering via Discrete Time Quantum Walk}
\author{
Junhong Nie
\thanks{State Key Lab of Processors, Institute of Computing Technology, Chinese Academy of Sciences.}
\thanks{School of Computer Science and Technology, University of Chinese Academy of Sciences.}
\\ \texttt{niejunhong19z@ict.ac.cn}
\and
Meng Li
\footnotemark[1]
\thanks{Corresponding author}
\\ \texttt{limeng2021@ict.ac.cn}
\and
Xiaoming Sun
\footnotemark[1]
\thanks{CAS Center for Excellence in Topological Quantum Computation, University of Chinese Academy of Sciences.}
\\ \texttt{sunxiaoming@ict.ac.cn}
}

\begin{document}
\maketitle

\begin{abstract}
    Quantum state engineering, namely the generation and control of arbitrary quantum states, is drawing more and more attention due to its wide applications in quantum information and computation. 
    However, there is no general method in theory, and the existing schemes also depend heavily on the selected experimental platform.
    In this manuscript, we give two schemes for the engineering task of arbitrary quantum state in $c$-partite $d$-dimensional system, both of which are based on discrete-time quantum walk with a $2^c$-dimensional time- and position-dependent coin. The first procedure is a $d$-step quantum walk where all the $d$ coins are non-identity, while the second procedure is an $O(d)$-step quantum walk where only $O(\log d)$ coins are non-identity. A concrete example of preparing generalized Bell states is given to demonstrate the first scheme we proposed. We also show how these schemes can be used to reduce the cost of long-distance quantum communication when the particles involved in the system are far away from each other. 
    Furthermore, the first scheme can be applied to give an alternative approach to the quantum state preparation problem which is one of the fundamental tasks of quantum information processing. We design circuits for quantum state preparation with the help of our quantum state engineering scheme that match the best current result in both size and depth of the circuit asymptotically.
\end{abstract}

\maketitle

\section{Introduction}\label{sec:introduction}
In quantum information and quantum computation, high-dimensional states can do multiple controls simultaneously \cite{wang2020qudits}, which may reduce the experimental cost and circuit size \cite{lu2020quantum}. Multipartite high-dimensional states, especially high-dimensional entanglement \cite{amico2008entanglement}, can be used as important sources in plenty of quantum information protocols, such as quantum teleportation \cite{bennett1993teleporting}, quantum key distribution \cite{durt2004security,walborn2006quantum,sheridan2010security} and quantum secret sharing \cite{keet2010quantum}. Thus, quantum state engineering, i.e., the realization of coherent dynamics or multipartite high-dimensional quantum states by manipulating appropriate quantum mechanical systems, has become a hot topic in modern physics and computer sciences.

Discrete time quantum walk acts as a powerful tool in a lot of areas of quantum information and computation. From the computational theory perspective, it is universal for quantum computation \cite{childs2009universal,childs2013universal,lovett2010universal,underwood2010universal}, and serves as an important part in various searching algorithms \cite{aaronson2003quantum,ambainis2007quantum,ambainis2020quadratic}. Apart from that, discrete-time quantum walk has been experimentally illustrated in some physical platforms, such as ion traps \cite{xue2009quantum,schmitz2009quantum}, optical \cite{chen2018observation,tang2018experimental}, and superconducting processor \cite{yan2019strongly,gong2021quantum}.

Due to the flexibility and experimental feasibility of quantum walk, it has also been applied to the preparation of high-dimensional states \cite{chandrashekar2008optimizing,majury2018robust}. In general, there are three kinds of quantum coins in quantum walks: time-dependent, position-dependent, time- and position-dependent. Innocent et.al propose a scheme for preparing most of the high-dimensional states on a single particle using a one-dimensional quantum walk with time-dependent coins \cite{innocenti2017quantum}, which is later experimentally realized in linear-optics platform \cite{giordani2019experimental} and further serves as an experimental example in Ref.\cite{suprano2021dynamical}. Moreover, Kadiri gives a sufficient condition of what states can be generated by quantum walk with time-dependent coin, of the whole system including both position and coin space \cite{kadiri2022steered}. On the other hand, Montero shows that quantum walks with time- and position-dependent coins can be used to retrieve any given distribution on a single particle \cite{montero2016classical,montero2017quantum}, and this theoretical result has been experimentally realized recently \cite{zhang2022arbitrary}.  

However, most of the current schemes are only for single qudit states, or there are limitations on the dimension of the bipartite quantum states discussed above.
Motivated by this, we focus on preparing arbitrary states in multipartite systems. Due to the essence of preparing multipartite high-dimensional entanglements and the variety of physical platforms, we aim to seek a general framework to accomplish this task by taking advantage of quantum walk with time- and position-dependent coins. Here, we first put forward two quantum walk procedures generating arbitrary $c$-partite $d$-dimensional quantum state $\ket{\phi}$ with the help of a $2^c$-dimension time- and position-dependent coin. The first one has exactly $d$ steps where all the $d$ coins are non-identity. The second scheme is an $O(d)$-step quantum walk where only $O(\log d)$ coins are non-identity. Both of these procedures has nearly optimal steps. 

We also illustrate the first scheme with two examples, hoping for encouraging potential applications of engineering multipartite high-dimensional quantum states. The first example is about preparing high-dimensional entangled states when the particles involved in the system are far away from each other. Utilizing a variation of our scheme with some standard quantum circuit techniques, one can prepare an arbitrary $d$-dimensional entangled state over a two-qudit system using $O(d^2)$ long-distance CNOT gates. This may have practical applications in quantum networks \cite{kimble2008quantum,wehner2018quantum} and distributed quantum computing \cite{broadbent2009universal,gottesman2012longer,komar2014quantum}. Another example is the relationship between quantum engineering and quantum state preparation. Quantum state preparation is a fundamental problem asking for a quantum circuit preparing any given $n$-qubit quantum state of size and depth as small as possible \cite{grover2002creating,bergholm2005quantum,plesch2011quantum,zhang2021low,sun2021asymptotically}. When the number of ancillary qubits is $O(n)$, the asymptotically optimal circuit accomplishing this task has size $\Theta(2^n)$ and depth $\Theta(\frac{2^n}{n})$ \cite{sun2021asymptotically}. We show that the same bound can also be reached simultaneously by treating the problem as engineering a bipartite $2^\frac{n}{2}$-dimensional quantum state and utilizing our scheme.

This paper is organized as follows. In \autoref{sec:preliminaries}, we briefly introduce the model of quantum walks we'll base on. In \autoref{sec:scheme}, we propose our first scheme to engineer an arbitrary multipartite high-dimensional quantum state based on quantum walk with time- and position-dependent coins. The bipartite case where $c=2$ is detailed in \autoref{ssec:bipartite} and the general multipartite case is sketched in \autoref{ssec:multipartite}. In \autoref{sec:alt_scheme}, we show our second scheme which has $O(d)$ steps while only $O(\log d)$ coins are non-identity. We only detail the bipartite case with $d$ a power of $2$. After that in \autoref{sec:application}, we give three examples where our scheme may be used. The first one is an illustration of our scheme by taking generalized Bell state as an example, which is shown in \autoref{ssec:Bell}. In \autoref{ssec:distributed}, a modified version is given aiming to reduce the cost of distant quantum communication when the system is ``distributed''. And the relationship between quantum state engineering and quantum state preparation is presented in \autoref{ssec:QSP}. Finally, we end with a summary and outlook in \autoref{sec:discussion}.

\section{Preliminaries}\label{sec:preliminaries}
Quantum walks are analogies of the well-known random walks in which at each step one tosses some coins (maybe biased) to decide which direction to go. In this work, the system where quantum walks apply to consists of $c$ particles and a $2^c$-dimension coin. That is, the system is the product of two Hilbert spaces: position space $\mathcal{H}_P$ in which the particles live and coin space $\mathcal{H}_C$. We use computational basis, i.e. $\mathcal{H}_P$ spanned by $\{\ket{\mathbf{x}}:\mathbf{x}\in \mathbb{Z}^c,\mathbf{x}\ge 0\}$, and $\mathcal{H}_C$ spanned by $\{\ket{\mathbf{c}}:\mathbf{c}\in\{0,1\}^c\}$. The initial state of the system is $\ket{0}^{\otimes c}\otimes\ket{0}^{\otimes c}$.

The $k$-th step of a quantum walk on $\mathcal{H}=\mathcal{H}_P\otimes\mathcal{H}_C$ is described as a unitary $W^{(k)}=SC^{(k)}$, where $C^{(k)}$ is the coin flipping operator and $S$ is the conditional shift operator. Generally speaking, there are three kinds of coin operators which have also been illustrated experimentally in the literature, time-dependent \cite{banuls2006quantum,xue2015experimental}, position-dependent \cite{suzuki2016asymptotic,ahmad2020one}, both time- and position-dependent \cite{kurzynski2013quantum,hou2018deterministic}. 
As mentioned in \autoref{sec:introduction}, in this work we use both time- and position-dependent coin flipping operator, that is:
\[C^{(k)}=\sum_{\mathbf{x}\ge 0}\ket{\mathbf{x}}\bra{\mathbf{x}}\otimes C^{(k)}(\mathbf{x}).\]
The conditional shift operator $S$ is defined as
\[S=\sum_{\substack{\mathbf{x}\ge 0\\ \mathbf{c}\in\{0,1\}^c}}\ket{\mathbf{x}+\mathbf{c}}\bra{\mathbf{x}}\otimes\ket{\mathbf{c}}\bra{\mathbf{c}}.\]

Let's illustrate this with the special case $c=2$. 
The initial state is $\ket{0,0}\otimes\ket{0,0}$. For this four dimensional coin, we may use alternative notations
\begin{align*}
    &\ket{\circlearrowleft}=\ket{0,0},
    \ket{\rightarrow}=\ket{1,0},\\
    &\ket{\uparrow}=\ket{0,1},
    \ket{\nearrow}=\ket{1,1}
\end{align*}
to describe the computational basis on 2D-grid, so the initial state can also be written $\ket{0,0}\otimes\ket{\circlearrowleft}$. Now we can define the coin operator $C^{(k)}$ and conditional shift operator $S$:
\begin{equation}\label{eq:C}
    C^{(k)}=\sum_{x,y\ge 0}\ket{x,y}\bra{x,y}\otimes C^{(k)}(x,y),
\end{equation}
where $C^{(k)}(x,y)$'s are all $4\times 4$ unitaries, and
\begin{equation}\label{eq:S}
    \begin{aligned}
        S(\ket{x,y}\otimes\ket{\circlearrowleft})&=\ket{x,y}\otimes\ket{\circlearrowleft},\\
        S(\ket{x,y}\otimes\ket{\rightarrow})&=\ket{x+1,y}\otimes\ket{\rightarrow},\\
        S(\ket{x,y}\otimes\ket{\uparrow})&=\ket{x,y+1}\otimes\ket{\uparrow},\\
        S(\ket{x,y}\otimes\ket{\nearrow})&=\ket{x+1,y+1}\otimes\ket{\nearrow}.
    \end{aligned}
\end{equation}
In fact, this direction calibration is equivalent to the usual method (the particle moves left, right, up or down according to the coin state), and in this way we can have a much cleaner and more intuitive representation.

\section{Theoretical scheme for engineering quantum states}\label{sec:scheme}
One can apparently make use of the scheme from Ref. \cite{innocenti2017quantum} on $c$ particles to produce arbitrary product states with only time-dependent coins used. However, engineering of multipartite entanglement states using only time-dependent coins is a bit more tricky. Indeed, one can encode the basis of the whole system in a zigzag fashion and utilize the scheme on the one-dimensional line, but this procedure takes $d^c$ steps of quantum walk, and needs us to design coin operators delicately (and $O(d^c)$ reaches the lower bound using time-dependent coins too, which can be proved by a counting argument). Therefore, these schemes are not ideal from the perspective of the number of steps.

In this section, we will elaborate the theoretical protocol of realizing arbitrary multipartite high-dimensional quantum state engineering via a $d$-step quantum walk, which is nearly optimal. Here we fully explore the potential of the quantum walk with time- and position-dependent coins by focusing on the critical positions and splitting coin operators.

\subsection{Bipartite case}\label{ssec:bipartite}
Now we explain how our method works in bipartite case. We make use of quantum walks with $c=2$, which is detailed in \autoref{sec:preliminaries}. Say one has an arbitrary state to be prepared $\ket{\phi}=\sum_{x,y=0}^{d-1}\alpha_{x,y}\ket{x,y}$, and the initial state is $\ket{\Psi^{(0)}}=\ket{0,0}\otimes\ket{\circlearrowleft}$, then the procedure is described as the following $d$ steps:
\begin{equation}\label{eq:preparation}
    \ket{\phi}\otimes\ket{\circlearrowleft}=W^{(d-1)}W^{(d-2)}\cdots W^{(1)}W^{(0)}\ket{\Psi^{(0)}}.
\end{equation}
The design of coin operators is the key to make the procedure suitable for arbitrary target quantum states. This may seem as a strong requirement, but later we shall see it's not the case: most of $C^{(k)}(x,y)$'s are identity.

Instead of directly design the coin operators in the quantum walk procedure, we accomplish the engineering task via an alternative procedure and turn it into the standard quantum walk procedure described in \autoref{eq:preparation}. The procedure we are looking at is denoted as
\[V^{(d-2)}\cdots V^{(1)}V^{(0)}\ket{\Psi^{(0)}}.\]
Later we'll see the definition of the operators $V^{(k)}$. Also, we define some intermediate states:
\begin{align*}
    \ket{\Psi^{(1)}}&=V^{(0)}\ket{\Psi^{(0)}},\\
    \ket{\Psi^{(2)}}&=V^{(1)}\ket{\Psi^{(1)}},\\
    &\dots,\\
    \ket{\Psi^{(d-1)}}&=V^{(d-2)}\ket{\Psi^{(d-2)}}.
\end{align*}
For these $\ket{\Psi^{(k)}}$'s, we make a crucial convention that their coin spaces are all in the particular state $\ket{\circlearrowleft}$. That is, they're all of the form
\[\ket{\Psi^{(k)}}=\ket{\phi^{(k)}}\otimes\ket{\circlearrowleft},\]
in which $\ket{{\phi^{(k)}}}=\sum_{x,y=0}^{k}\alpha^{(k)}_{x,y}\ket{x,y}$ is spanned by $\{\ket{l,r}:0\le l,r\le k\}$ for all $k=0,1,\dots,d-1$ and $\ket{\phi^{(d-1)}}=\ket{\phi}$. From now on, we use the notation $\ket{\Psi^{(k)}}$ and $\ket{\phi^{(k)}}$ interchangeably.

Now we define the operators $V^{(k)}$. Each $C^{(k)}$ in \autoref{eq:preparation} is decomposed into two coin operators: $C^{(k)}=C_1^{(k)}C_2^{(k)}$ for $k=0,\dots,d-1$ with $C_2^{(0)}=C_1^{(d-1)}=I$. And we define $V^{(k)}=C_2^{(k+1)}SC_1^{(k)}$ for all $k=0,1,\dots,d-2$. Intuitively, the coin operator $C_1^{(k)}$'s fork the directions according to the $\alpha^{(k+1)}_{x,y}$'s, and $C_2^{(k)}$'s are for the purpose of turning the states of coin space into $\ket{\circlearrowleft}$. Thus as long as the state after shift operator ($SC_1^{(k)}\ket{\Psi^{(k)}}$) is given, the operator $C_2^{(k+1)}$ is settled automatically.

With these definitions in hand, we show how to engineer the state $\ket{\phi}$. This is described in an inductive fashion: suppose for any $\ket{\Psi^{(k)}}$, there exists operators $V^{(0)},V^{(1)},\dots,V^{(k-1)}$ such that
\[\ket{\Psi^{(k)}}=V^{(k-1)}\cdots V^{(1)}V^{(0)}\ket{\Psi^{(0)}},\]
we claim that for any given state $\ket{\phi^{(k+1)}}$, there exists operators $V^{(0)},V^{(1)},\dots,V^{(k)}$ such that
\[\ket{\Psi^{(k+1)}}=V^{(k)}\cdots V^{(1)}V^{(0)}\ket{\Psi^{(0)}}.\]
Once this claim is proved, we can indeed reach our goal: simply let $\ket{\Psi^{(d-1)}}$ be equal to $\ket{\phi}\otimes\ket{\circlearrowleft}$, then there exists $V^{(0)},V^{(1)},\dots,V^{(d-2)}$ such that
\[\ket{\Psi^{(d-1)}}=V^{(d-2)}\cdots V^{(1)}V^{(0)}\ket{\Psi^{(0)}}.\]
This indeed corresponds to a quantum walk procedure described in \autoref{eq:preparation}, because $S\ket{\Psi^{(d-1)}}=\ket{\Psi^{(d-1)}}$.

Given $\ket{\phi^{(k+1)}}$, to prove the claim, we delicately pick some $\ket{\phi^{(k)}}$ and computes two coins $C_1^{(k)}$ and $C_2^{(k+1)}$, such that $\ket{\Psi^{(k+1)}}=C_2^{(k+1)}SC_1^{(k)}\ket{\Psi^{(k)}}$ for all $k=0,1,\dots,d-2$. Since the coin operator can take full control of the position space, let's denote
\begin{align*}
    C_i^{(k)}=\sum_{x,y\ge 0}\ket{x,y}\bra{x,y}\otimes C_i^{(k)}(x,y),
\end{align*}
where $i=1,2$. Note that $\ket{{\phi^{(k)}}}$ is spanned by $\{\ket{l,r}:0\le l,r\le k\}$, which means $C_1^{(k)}(x,y)$ is meaningless for $x>k$ or $y>k$, and thus they can be set identity. To make the coin operator as simple as possible, $C_1^{(k)}(x,y)$ for $x,y<k$ are set identity too. This in fact leads to the state $\ket{\phi^{(k)}}$ we choose. 

The base case $k=0$ is simple. The initial state $\ket{\Psi^{(0)}}$ is fixed, and it's straightforward to verify the following $C_1^{(0)}$ and $C_2^{(1)}$ makes $\ket{\Psi^{(1)}}=C_2^{(1)}SC_1^{(0)}\ket{\Psi^{(0)}}$, where $C_1^{(0)}$ satisfies
\[C_1^{(0)}(0,0)\ket{\circlearrowleft}=\alpha^{(1)}_{0,0}\ket{\circlearrowleft}+\alpha^{(1)}_{1,0}\ket{\rightarrow}+\alpha^{(1)}_{0,1}\ket{\uparrow}+\alpha^{(1)}_{1,1}\ket{\nearrow},\]
and
\[C_2^{(1)}(x,y)=\begin{cases}
    I\otimes I,&x=0,y=0;\\
    X\otimes I,&x=1,y=0;\\
    I\otimes X,&x=0,y=1;\\
    X\otimes X,&x=1,y=1.
\end{cases}\]

For $k>0$, suppose the given state is $\ket{\phi^{(k+1)}}=\sum_{x,y=0}^{k+1}\alpha^{(k+1)}_{x,y}\ket{x,y}$, then the algorithm sets
\begin{equation}
    \label{eq:alpha}
    \alpha^{(k)}_{x,y}=\begin{cases}
    \alpha^{(k+1)}_{x,y},&x,y<k;\\
    \sqrt{|\alpha^{(k+1)}_{x,y}|^2+|\alpha^{(k+1)}_{x+1,y}|^2},&x=k,y<k;\\
    \sqrt{|\alpha^{(k+1)}_{x,y}|^2+|\alpha^{(k+1)}_{x,y+1}|^2},&x<k,y=k;\\
    \sqrt{|\alpha^{(k+1)}_{x,y}|^2+|\alpha^{(k+1)}_{x+1,y}|^2+|\alpha^{(k+1)}_{x,y+1}|^2+|\alpha^{(k+1)}_{x+1,y+1}|^2},&x=y=k.
    \end{cases}
\end{equation}

Now suppose this particular $\ket{\Psi^{(k)}}$ has been prepared, the operators $C_1^{(k)}(x,y)$ are given by the following formula. Note that for simplicity, we slightly abuse the notations such that the whole term is zero while dividing by zero.
\begin{equation}
    \label{eq:C1}
    C_1^{(k)}(x,y)\ket{\circlearrowleft}=\begin{cases}
    \ket{\circlearrowleft},&x,y<k;\\
    \left(\alpha^{(k+1)}_{x,y}\ket{\circlearrowleft}+\alpha^{(k+1)}_{x+1,y}\ket{\rightarrow}\right)/\alpha^{(k)}_{x,y},&x=k,y<k;\\
    \left(\alpha^{(k+1)}_{x,y}\ket{\circlearrowleft}+\alpha^{(k+1)}_{x,y+1}\ket{\uparrow}\right)/\alpha^{(k)}_{x,y},&x<k,y=k;\\
    \left(\alpha^{(k+1)}_{x,y}\ket{\circlearrowleft}+\alpha^{(k+1)}_{x+1,y}\ket{\rightarrow}+\alpha^{(k+1)}_{x,y+1}\ket{\uparrow}+\alpha^{(k+1)}_{x+1,y+1}\ket{\nearrow}\right)/\alpha^{(k)}_{x,y},&x=y=k,
    \end{cases}
\end{equation}
And $C_2^{(k)}(x,y)$ satisfies
\begin{equation}\label{eq:C2}
    \begin{aligned}
        &C_2^{(k)}(x,y)\ket{\circlearrowleft}=\ket{\circlearrowleft},&x,y<k;\\
        &C_2^{(k)}(x,y)\ket{\rightarrow}=\ket{\circlearrowleft},&x=k,y<k;\\
        &C_2^{(k)}(x,y)\ket{\uparrow}=\ket{\circlearrowleft},&x<k,y=k;\\
        &C_2^{(k)}(x,y)\ket{\nearrow}=\ket{\circlearrowleft},&x=y=k.
    \end{aligned}
\end{equation}
It's straightforward to see these $C_1^{(k)}(x,y)$'s and $C_2^{(k)}(x,y)$'s can be set to unitaries and they can indeed achieve our goal. Let's take the case $x=k,y<k$ as example, and the proof of the remaining cases are similar. The initial state at $x=k,y<k$ is $\alpha^{(k)}_{x,y}\ket{x,y}\otimes\ket{\circlearrowleft}$, and after $C_1^{(k)}(x,y)$, the shift operator $S$ and $C_2^{(k+1)}(x,y)$, the state becomes
\begin{align*}
    &\alpha^{(k)}_{x,y}\ket{x,y}\otimes\ket{\circlearrowleft}\\
    \mapsto&\alpha^{(k)}_{x,y}\ket{x,y}\otimes\left(\alpha^{(k+1)}_{x,y}\ket{\circlearrowleft}+\alpha^{(k+1)}_{x+1,y}\ket{\rightarrow}\right)/\alpha^{(k)}_{x,y}\\
    \mapsto&\alpha^{(k+1)}_{x,y}\ket{x,y}\otimes\ket{\circlearrowleft}+\alpha^{(k+1)}_{x+1,y}\ket{x+1,y}\otimes\ket{\rightarrow}\\
    \mapsto&\left(\alpha^{(k+1)}_{x,y}\ket{x,y}+\alpha^{(k+1)}_{x+1,y}\ket{x+1,y}\right)\otimes\ket{\circlearrowleft}.
\end{align*}
Also note that the $C_1^{(k)}(x,y)$'s and $C_2^{(k)}(x,y)$'s can be set identity for all $x,y<k$.

In fact, the explicit form of the $\alpha^{(k)}_{x,y}$'s in our algorithm for every step $k<d$ can be directly expressed by the amplitudes of the desired state $\ket{\phi}$, i.e. $\alpha_{x,y}$'s. For a fixed $k$, we have the following observations: $\alpha^{(k)}_{x,y}$'s are zero except for the case $x\le k,y\le k$. In addition, $\alpha^{(k)}_{x,y}=\alpha_{x,y}$ for $x<k,y<k$. So the only unsettled cases are $x=k,y<k$; $x<k,y=k$; and $x=k,y=k$. It's straightforward to see that
\begin{equation}
    \label{eq:alpha_ex}
    \alpha^{(k)}_{x,y}=\begin{cases}
    \alpha_{x,y},&x,y<k;\\
    \sqrt{\sum_{z=k}^{d-1}|\alpha_{z,y}|^2},&x=k,y<k;\\
    \sqrt{\sum_{w=k}^{d-1}|\alpha_{x,w}|^2},&x<k,y=k;\\
    \sqrt{\sum_{z,w=k}^{d-1}|\alpha_{z,w}|^2},&x=y=k.
    \end{cases}
\end{equation}
One can prove this by plugging it into \autoref{eq:alpha} and making induction on $k$. With the help of \autoref{eq:alpha_ex}, one can construct the coin operators $C_1^{(k)}(x,y)$ for every $k$ explicitly.

To compute the elements of $C^{(k)}(x,y)$'s, one needs to compute the coefficients $\alpha_{x,y}^{(k)}$ for every $k$ first, according to \autoref{eq:alpha}. Notice that at step $k$, there are only $2k-1$ of these coefficients that differs from $\alpha_{x,y}$, so the total effort of computing these coefficients for all $k$ is $O(d^2)$. After that, the elements of the coin operators can be computed according to \autoref{eq:C1} and \autoref{eq:C2}. Again, recall that at step $k$, only $2k-1$ out of $k^2$ of the operators $C^{(k)}_{x,y}$ are non-identity, and they are all $4\times 4$ matrices whose elements can be computed from the $\alpha_{x,y}^{(k)}$'s. Thus, the total effort of computing these operators is $O(d^2)$, too. In summary, this algorithm takes $O(d^2)$ time to compute all the coin operators used in engineering $\ket{\phi}$.

\subsection{Multipartite case}\label{ssec:multipartite}
Multipartite case is a natural generalization of the bipartite case, so here we briefly sketch the scheme. Say again one has an arbitrary state to be prepared $\ket{\phi}=\sum_{\mathbf{x}< d}\alpha_{\mathbf{x}}\ket{\mathbf{x}}$, and the initial state is $\ket{\Psi^{(0)}}=\ket{0^c}\otimes\ket{0^c}$. Then the quantum walk procedure is
\[\ket{\phi}\otimes\ket{0^c}=SC^{(d-1)}SC^{(d-2)}\cdots SC^{(1)}SC^{(0)}\ket{\Psi^{(0)}}.\]

Again, we look at the alternative procedure
\[V^{(d-2)}\cdots V^{(1)}V^{(0)}\ket{\Psi^{(0)}}.\]
Denote $\ket{\Psi^{(k)}}=\ket{\phi^{(k)}}\otimes\ket{0^c}$ the intermediate state of step $k$ in which \[\ket{\phi^{(k)}}=\sum_{\mathbf{x}\le k}\alpha^{(k)}_{\mathbf{x}}\ket{\mathbf{x}},\]
and
\[\ket{\Psi^{(k+1)}}=V^{(k)}\ket{\Psi^{(k)}}\]
for all $k=0,1,\dots,d-2$. Remind that $V^{(k)}=C_2^{(k+1)}SC_1^{(k)}$, 
$C_i^{(k)}=\sum_{0\le\mathbf{x}\le k}\ket{\mathbf{x}}\bra{\mathbf{x}}\otimes C_i^{(k)}(\mathbf{x})$, where $i=1,2$.

Now we describe how to design the operators $V^{(k)}$. We achieve this in an inductive fashion: suppose for any $\ket{\Psi^{(k)}}$, there exists an alternative procedure of $k-1$ steps, we show that for any given state $\ket{\phi^{(k+1)}}$, there exists an alternative procedure of $k$ steps. That is, given $\ket{\phi^{(k+1)}}$, we delicately pick some $\ket{\phi^{(k)}}$ and computes two coins $C_1^{(k)}$ and $C_2^{(k+1)}$, such that $\ket{\Psi^{(k+1)}}=C_2^{(k+1)}SC_1^{(k)}\ket{\Psi^{(k)}}$.

The base case $k=0$ is simple and similar to the bipartite case in \autoref{ssec:bipartite}. For $k>0$, given arbitrary state $\ket{\phi^{(k+1)}}=\sum_{\mathbf{x}\le k+1}\alpha^{(k+1)}_{\mathbf{x}}\ket{\mathbf{x}}$, the algorithm sets $\alpha^{(k)}_{\mathbf{x}}$'s in the following way. For $\mathbf{x}<k$, $\alpha_{\mathbf{x}}^{(k)}$ is simply set to be equal to $\alpha_{\mathbf{x}}^{(k+1)}$. For $\mathbf{x}\le k$ with $x_i=k$ for some $i$, define a vector cylinder $V_{\mathbf{x}}=\{\mathbf{z}\in\{0,1\}^c:z_j=0\text{ if }x_j<k\}$, then the algorithm sets
\[\alpha^{(k)}_{\mathbf{x}}=\sqrt{\sum_{\mathbf{z}\in V_{\mathbf{x}}}\left|\alpha^{(k+1)}_{\mathbf{x}+\mathbf{z}}\right|^2}. \]

Now suppose the state $\ket{\phi^{(k)}}$ defined above is prepared, $C_1^{(k)}(\mathbf{x})$ is given by
\[C_1^{(k)}(\mathbf{x})\ket{0^c}=\sum_{\mathbf{z}\in V_{\mathbf{x}}}\alpha^{(k+1)}_{\mathbf{x}+\mathbf{z}}\ket{\mathbf{z}}/\alpha^{(k)}_{\mathbf{x}}.\]
And $C_2^{(k)}(\mathbf{x})$ satisfies
\[C_2^{(k)}(\mathbf{x})\ket{\mathbf{z}}=\ket{0^c}\]
for all $\mathbf{z}\in V_{\mathbf{x}}$. Note again that the $C_1^{(k)}(\mathbf{x})$'s and $C_2^{(k)}(\mathbf{x})$'s can be set identity for all $\mathbf{x}<k$.

\section{An alternative scheme for engineering quantum states}\label{sec:alt_scheme}
In this section, we elaborate another theoretical protocol of realizing arbitrary multipartite high-dimensional quantum state engineering via a $O(d)$-step quantum walk. While the scheme described in \autoref{sec:scheme} consists of $d$ steps of quantum walk that each coin operator is non-identity, this alternative scheme has only $O(\log d)$ non-identity coin operators in total. This is achieved by taking advantage of time- and position-dependent coins further such that the engineering procedure becomes much more paralleled. We will illustrate this scheme by only the bipartite case when the dimension $d$ is a power of $2$, scheme for general $d$ and the multipartite case are both natural generalizations that are omitted.

Below we adopt the settings and notions used in \autoref{ssec:bipartite}. Suppose the state to be prepared is $\ket{\phi}=\sum_{x,y=0}^{d-1}\alpha_{x,y}\ket{x,y}$, and the initial state is $\ket{\Psi^{(0)}}=\ket{0,0}\otimes\ket{\circlearrowleft}$. The whole procedure is described as \autoref{eq:preparation}. Again, instead of directly design the coin operators, we decompose each $C^{(k)}$ into $C^{(k)}=C_1^{(k)}C_2^{(k)}$ and seek for another procedure that is equivalent to the original one. This time we look at a slightly different procedure, which is denoted as
\[V^{(\log d-1)}\dots V^{(1)}V^{(0)}\ket{\Psi^{(0)}},\]
where $V^{(i)}=C_2^{(i+1)}S^\frac{d}{2^{i+1}}C_1^{(i)}$ for $i=0,1,\dots,\log d-1$. One can verify that this procedure is indeed a quantum walk of $O(d)$ steps.

Intuitively, this scheme builds up the whole state by first engineering a state that has non-zero amplitudes only on some particular positions, say $\ket{0,0},\ket{\frac{d}{2},0},\ket{0,\frac{d}{2}},\ket{\frac{d}{2},\frac{d}{2}}$, and assuming one can engineer arbitrary bipartite states with dimension $\frac{d}{2}$, then the whole state can be engineered by starting from these four positions parallel at each of which calls the state engineering procedure for dimension $\frac{d}{2}$.

Like \autoref{ssec:bipartite}, we describe this scheme in an inductive fashion too. Denote $\ket{\phi^{(k)}}=\sum_{x,y=0}^{2^k-1}\alpha_{x,y}^{(k)}\ket{x,y}$ be an arbitrary state and $\ket{\Psi^{(k)}}=\ket{\phi^{(k)}}\otimes\ket{\circlearrowleft}$, for $k=0,1,\dots,\log d-1$, and let $\ket{\phi^{(\log d)}}=\ket{\phi}$. We claim that suppose for any $\ket{\Psi^{(k)}}$, there exists operators $V^{(0)},V^{(1)},\dots,V^{(k-1)}$ of the form $V^{(j)}=C_2^{(j+1)}S^{2^{k-1-j}}C_1^{(j)}$ such that
\[\ket{\Psi^{(k)}}=V^{(k-1)}\cdots V^{(1)}V^{(0)}\ket{\Psi^{(0)}},\]
then for any given state $\ket{\phi^{(k+1)}}$, there exists operators $U^{(0)},U^{(1)},\dots,U^{(k)}$ of the form $U^{(j)}=C_2^{(j+1)}S^{2^{k-j}}C_1^{(j)}$ such that
\[\ket{\Psi^{(k+1)}}=U^{(k)}\cdots U^{(1)}U^{(0)}\ket{\Psi^{(0)}}.\]
Once this claim is proved, we can indeed reach our goal: simply let $\ket{\Psi^{(\log d)}}$ be equal to $\ket{\phi}\otimes\ket{\circlearrowleft}$, then there exists $V^{(0)},V^{(1)},\dots,V^{(\log d-1)}$ such that
\[\ket{\Psi^{(\log d)}}=V^{(\log d-1)}\cdots V^{(1)}V^{(0)}\ket{\Psi^{(0)}}.\]

The base case $k=1$ is identical with the base case in \autoref{ssec:bipartite}. For $k>1$, suppose the given state is $\ket{\phi^{(k+1)}}=\sum_{x,y=0}^{2^{k+1}-1}\alpha^{(k+1)}_{x,y}\ket{x,y}$, we construct four states starting at the four positions mentioned earlier. Denote
\begin{align*}
    \ket{\phi^{(k)}_{0,0}}&=\frac{1}{\sqrt{\gamma_{0,0}}}\sum_{x,y=0}^{2^k-1}\alpha^{(k+1)}_{x,y}\ket{x,y},
    &\gamma_{0,0}=\sum_{x,y=0}^{2^k-1}\left|\alpha^{(k+1)}_{x,y}\right|^2;\\
    \ket{\phi^{(k)}_{1,0}}&=\frac{1}{\sqrt{\gamma_{1,0}}}\sum_{x,y=0}^{2^k-1}\alpha^{(k+1)}_{x+2^k,y}\ket{x,y},
    &\gamma_{1,0}=\sum_{x=2^k}^{2^{k+1}-1}\sum_{x,y=0}^{2^k-1}\left|\alpha^{(k+1)}_{x,y}\right|^2;\\
    \ket{\phi^{(k)}_{0,1}}&=\frac{1}{\sqrt{\gamma_{0,1}}}\sum_{x,y=0}^{2^k-1}\alpha^{(k+1)}_{x,y+2^k}\ket{x,y},
    &\gamma_{0,1}=\sum_{x=0}^{2^k-1}\sum_{y=2^k}^{2^{k+1}-1}\left|\alpha^{(k+1)}_{x,y}\right|^2;\\
    \ket{\phi^{(k)}_{1,1}}&=\frac{1}{\sqrt{\gamma_{1,1}}}\sum_{x,y=0}^{2^k-1}\alpha^{(k+1)}_{x+2^k,y+2^k}\ket{x,y},
    &\gamma_{1,1}=\sum_{x,y=2^k}^{2^{k+1}-1}\left|\alpha^{(k+1)}_{x,y}\right|^2.
\end{align*}
According to the assumption, each of these four states can be engineered by a $k$-step procedure which, for the procedure engineering $\ket{\phi^{(k)}_{a,b}}$ we denote $U^{(k+1)}_{a,b}\cdots U^{(2)}_{a,b}U^{(1)}_{a,b}$, for $a,b\in\{0,1\}$ respectively, where $U^{(j)}_{a,b}=C_{2,a,b}^{(j+1)}S^{2^{k-j}}C_{1,a,b}^{(j)}$ by the definition of these operators.

Now we show the design of the operators $U^{(j)}$ for $j=0,1,\dots, k$. The operator $U^{(0)}=C_2^{(1)}S^{2^k}C_1^{(0)}$ is described as follows:
\[
    C_1^{(0)}(0,0)\ket{\circlearrowleft}=\sqrt{\gamma_{0,0}}\ket{\circlearrowleft}+\sqrt{\gamma_{1,0}}\ket{\rightarrow}+\sqrt{\gamma_{0,1}}\ket{\uparrow}+\sqrt{\gamma_{1,1}}\ket{\nearrow}
\]
and $C_1^{(0)}(x,y)=I$ for $x>0$ or $y>0$. And $C_2^{(k)}(x,y)$ satisfies
\begin{align*}
        &C_2^{(1)}(x,y)\ket{\circlearrowleft}=\ket{\circlearrowleft},&x,y=0;\\
        &C_2^{(1)}(x,y)\ket{\rightarrow}=\ket{\circlearrowleft},&x=2^k,y=0;\\
        &C_2^{(1)}(x,y)\ket{\uparrow}=\ket{\circlearrowleft},&x=0,y=2^k;\\
        &C_2^{(1)}(x,y)\ket{\nearrow}=\ket{\circlearrowleft},&x=y=2^k;
\end{align*}
and $C_2^{(1)}(x,y)=I$ for other positions. It is easily checked that after applying operator $U^{(0)}$ on the initial state, we get the following state:
\[\left(\sqrt{\gamma_{0,0}}\ket{0,0}+\sqrt{\gamma_{1,0}}\ket{2^k,0}+\sqrt{\gamma_{0,1}}\ket{0,2^k}+\sqrt{\gamma_{1,1}}\ket{2^k,2^k}\right)\otimes\ket{\circlearrowleft}.\]
The operator $U^{(j)}=C_2^{(j+1)}S^{2^{k-j}}C_1^{(j)}$ for $j=1,2,\dots k$ satisfies
\[C_1^{(j)}=\sum_{x,y=0}^{2^{k+1}-1}\ket{x,y}\otimes C_{1,\mathbf{1}_{x\ge 2^k},\mathbf{1}_{x\ge 2^k}}^{(j)}\]
and
\[C_2^{(j)}=\sum_{x,y=0}^{2^{k+1}-1}\ket{x,y}\otimes C_{2,\mathbf{1}_{x\ge 2^k},\mathbf{1}_{x\ge 2^k}}^{(j)}.\]
The correctness comes naturally from assumption about the $U^{(j)}_{a,b}$'s for $a,b\in\{0,1\}$, and the whole procedure has only $O(\log d)$ non-identity coin operators as promised.

\section{Applications of our scheme}\label{sec:application}
In this section, we show three applications of our scheme for engineering quantum states in multipartite systems. In \autoref{ssec:Bell}, we illustrate our scheme by taking engineering generalized Bell state as an example. In \autoref{ssec:distributed}, we show how to manipulate quantum engineering under distributed settings, where the particles in the system are far away from each other. We show $O(d^2)$ long-distance CNOT operations is enough for preparing bipartite $d$-dimensional states. Later in \autoref{ssec:QSP}, we show how the quantum walk procedure obtained by our scheme can be used to generate a quantum circuit of size $O(2^n)$ and depth $O(\frac{2^n}{n})$ preparing any $n$-qubit state with $O(n)$ ancillary qubits, which also reaches the best quantum state preparation method accomplishing this task.

\subsection{Taking generalized Bell state as example}\label{ssec:Bell}
Generalized Bell state, which is written as
\[\ket{\phi_{n,m}}=\frac{1}{\sqrt{d}}\sum_{j=0}^{d-1}e^{2\pi \mathrm{i}jn/d}\ket{j}\otimes\ket{(j+m)\bmod{d}}\]
for some $n,m$, firstly serves as basis for the measurements Alice used in the quantum teleportation scheme\cite{bennett1993teleporting}. Later in some quantum key distribution (QKD) protocols, it is used as the initial shared entanglements \cite{karimipour2002quantum}. Here, we illustrate our algorithm with this particular state.

We define the $\ket{\Psi^{(k)}}$'s and $\ket{\phi^{(k)}}$'s for $k=0,1,\dots,d-1$ as described in \autoref{ssec:bipartite}, and will give the coin operators $C_1^{(k)}$'s and $C_2^{(k)}$'s explicitly.

Let $\ket{\phi^{(d-1)}}=\ket{\phi_{n,m}}$, the algorithm sets
\[
    \alpha^{(d-1)}_{x,y}=\begin{cases}
    e^{2\pi \mathrm{i}jn/d}/\sqrt{d},&y\equiv x+m\pmod{d};\\
    0,&\text{otherwise}.
    \end{cases}
\]
For $k<d-1$, we suppose $0\le m<d$ without loss of generality. According to \autoref{eq:alpha_ex}, there are three non-trivial cases we need to consider.

Case 1: $x=k,y<k$. Following \autoref{eq:alpha_ex}, the only coordinates $(k,y)$ such that $\alpha^{(k)}_{k,y}$ isn't zero has its $y$ coordinate satisfying the following constraints:
\[\begin{cases}
    y\equiv z+m\pmod{d};\\
    k\le z<d.
\end{cases}\]
That is, there exists $t\in \mathbb{Z}$ that $k\le y-m-td<d$, which is
\[\frac{y-m}{d}-1<t\le\frac{y-m}{d}-\frac{k}{d}.\]
Notice that $0\le y<k$ and $0\le m<d$, so $-1<\frac{y-m}{d}<1$, hence the only possible value of $t$ is $0$ and $-1$. If $t=0$, then $y-m\ge k$, which contradicts $y<k$; otherwise $t=-1$, and the constraint $k+m-d\le y<m$ must be satisfied. 

Case 2: $x<k,y=k$. Similar to case 1, the only coordinates $(x,k)$ such that $\alpha^{(k)}_{x,k}$ isn't zero has its $x$ coordinate satisfying the following constraints:
\[\begin{cases}
    w\equiv x+m\pmod{d};\\
    k\le w<d.
\end{cases}\]
That is, there exists $t\in \mathbb{Z}$ that $k\le x+m+td<d$, which is
\[\frac{-x-m}{d}+\frac{k}{d}\le t<\frac{-x-m}{d}+1.\]
Notice again that $0\le x<k$ and $0\le m<d$, so $-2<\frac{-x-m}{d}\le 0$, hence $t$ can only be $0$ or $-1$. If $t=-1$, then $x\ge k-m+d$, which leads to a contradiction; otherwise $t=0$, and the constraint $k-m\le x<d-m$ must be satisfied.

Case 3: $x=k,y=k$. According to \autoref{eq:alpha_ex}, we need to count the number of integer solutions of the constraints below:
\[\begin{cases}
    w\equiv z+m\pmod{d};\\
    k\le z<d;\\
    k\le w<d.
\end{cases}\]
Fix a particular $z$ such that $k\le z<d$. Same as the analysis in case 2, there are exactly one integer $w$ such that $(z,w)$ satisfies these constraints when $k\le z<d-m$ or $k-m+d\le z<d$. When $d-m\le z<k-m+d$, there's no such $w$.
Define a function $\delta:\mathbb{R}\to\mathbb{R}$ as
\[\delta(z)=\begin{cases}
    1,&z\ge 0;\\
    0,&z<0.
\end{cases}\]
Denote the function $u:\mathbb{R}\to\mathbb{R}$ as $u(z)=z\delta(z)$.
Then the total number of integer solutions is $u(d-m-k)+u(m-k)$, which is denoted $\sigma_m(k)$.

To summarize, for $k<d$ we have
\[
    \label{eq:alpha_k_gb}
    \alpha^{(k)}_{x,y}=\begin{cases}
    \alpha_{x,y},&x,y<k;\\
    \frac{1}{\sqrt{d}},&x=k,k+m-d\le y<\min\{m,k\};\\
    \frac{1}{\sqrt{d}},&k-m\le x<\min\{d-m,k\},y=k;\\
    \sqrt{\frac{\sigma_m(k)}{d}},&x=y=k;\\
    0,&\text{otherwise}.
    \end{cases}
\]

One can obtain the corresponding coin operators via plugging the $\alpha^{(k)}_{x,y}$'s into \autoref{eq:C1}. Notice that for each $C_1^{(k)}(x,y)\in \mathbb{C}^{4\times 4}$, \autoref{eq:C1} has only four constraints, so he may pick the unitaries that are relatively simple. Here we give one possible version of the coin operators $C_1^{(k)}(x,y)$ for $k<d$ of preparing generalized Bell states directly:
\[
    C_1^{(k)}(x,y)=\begin{cases}
    e^{2\pi \mathrm{i}kn/d}I\otimes I,&x=k,y=k+m-d;\\
    X\otimes I,&x=k,k+m-d<y<\min\{m,k\};\\
    e^{2\pi \mathrm{i}(k-m)n/d}I\otimes I,&x=k-m,y=k;\\
    I\otimes X,&k-m<x<\min\{d-m,k\},y=k;\\
    D^{(k)},&x=y=k;\\
    I\otimes I,&\text{otherwise},
    \end{cases}
\]
in which $I,X,Y$ are Pauli matrices. 
And $D^{(k)}$ satisfies
\[D^{(k)}\ket{\circlearrowleft}=\begin{cases}
    \sqrt{\frac{1}{d-k}}e^{2\pi \mathrm{i}kn/d}\ket{\circlearrowleft}+\sqrt{1-\frac{1}{d-k}}\ket{\nearrow},&m=0;\\
    \sqrt{\frac{\delta(m-k-1)}{\sigma_m(k)}}\ket{\rightarrow}+\sqrt{\frac{\delta(d-m-k-1)}{\sigma_m(k)}}\ket{\uparrow}+\sqrt{\frac{\sigma_m(k+1)}{\sigma_m(k)}}\ket{\nearrow},&m\ne 0.
\end{cases}\]
The operator $C_2^{(k)}(x,y)$'s can be determined by \autoref{eq:C2}.

\subsection{Quantum engineering under distributed settings}\label{ssec:distributed}
In many quantum information and quantum communication protocols with entanglement involved, the particles in the system may be distant from each other \cite{kimble2008quantum,wehner2018quantum,broadbent2009universal,gottesman2012longer,komar2014quantum}. Consider a typical scene that several high-dimensional particles are distributed all around the world and some entangled state is required to be prepared on them. Although it is apparently impossible to accomplish this task without any manipulation of executing distant quantum operations, we aim to reduce the amount. To a certain extent, long-distance CNOT gate has been proved to be practical by several physical experiments \cite{daiss2021quantum,pompili2021realization}. Yet to the best of our knowledge, there is basically no physical experiments realizing distant gates on qudits.

Here, we will show that any state on a distributed quantum system can be effectively prepared in the sense of using as low the costs of distant quantum operations as possible. Specifically, suppose some people have $c$ particles of dimension $d$ distributed all over the world, and they want to prepare arbitrary state on these particles only using long-distance CNOT gates. In our quantum walk scheme as described in \autoref{ssec:distributed}, there are mainly two kinds of operators: the coin operators $C^{(k)}$ and the shift operator $S$. The shift operator turns out to be local, so we focus on the coin operators. Recall the form of $C^{(k)}$ shown in \autoref{eq:C}. They take full control of the position space and thus they are high-dimensional controlled operations. To avoid distant operations of this kind, we introduce several ancillary qubits for each part of the system. By delicately designed local operations, these long-distance operations are transformed into multi-qubit gates, which can be further decomposed into single-qubit gates and CNOT gates by standard techniques \cite{nielsen2002quantum}. This results in $O(d^c)$ long-distance CNOT gates.

\begin{figure*}[htbp]
    \centering
    \mbox{\Qcircuit @C=0.7em @R=1em{
        \lstick{\ket{0}} &\rstick{x}\qw\qwx[1] &\multigate{1}{S(a,b)} &\rstick{x}\qw\qwx[1] &\multigate{1}{S(a,b)} &\cds{3}{\cdots} &\rstick{x}\qw\qwx[1] &\multigate{1}{S(a,b)} &\qw\\
        \lstick{\ket{0}} &\rstick{y}\qw\qwx[1] &\ghost{S(a,b)} &\rstick{y}\qw\qwx[1] &\ghost{S(a,b)} &\qw &\rstick{y}\qw\qwx[1] &\ghost{S(a,b)} &\qw\\
        \lstick{\ket{0}} &\multigate{1}{C^{(0)}(x,y)} &\rstick{a}\qw\qwx[-1] &\multigate{1}{C^{(1)}(x,y)} &\rstick{a}\qw\qwx[-1] &\qw &\multigate{1}{C^{(d-1)}(x,y)} &\rstick{a}\qw\qwx[-1] &\qw\\
        \lstick{\ket{0}} &\ghost{C^{(0)}(x,y)} &\rstick{b}\qw\qwx[-1] &\ghost{C^{(1)}(x,y)} &\rstick{b}\qw\qwx[-1] &\qw &\ghost{C^{(d-1)}(x,y)} &\rstick{b}\qw\qwx[-1] &\qw
        \gategroup{1}{2}{4}{3}{0.7em}{--}
    }}
    \caption{Circuit of our algorithm. The controlled gate with label on its controlled space represents a ``uniformly controlled gate". That is, when the state of the position space is $\ket{x,y}$, $C^{(k)}(x,y)$ is executed on the coin space.}
    \label{fig:whole_circuit}
\end{figure*}
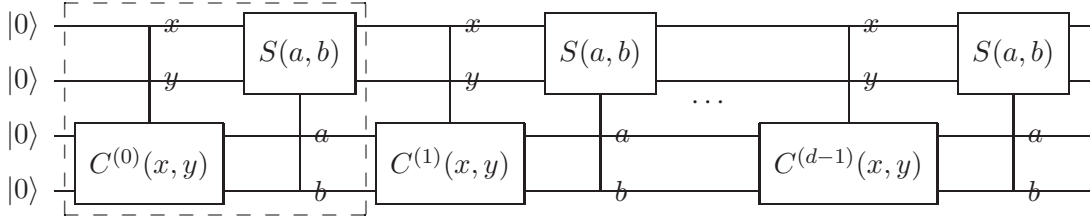

Now Let's explain how our scheme works under this setting in detial. We'll only involve bipartite case, and the multipartite case is a natural generalization. Suppose we have two particles belonging to two different sides of the system, and each side owns one qubit of the quantum coin. Formally, the Hilbert space we're in, namely $\mathcal{H}_P\otimes\mathcal{H}_C$, is isomorphic to $(\mathbb{C}^d\otimes\mathbb{C}^2)^{\otimes 2}$. To make it clear, we illustrate our procedure under quantum circuit model. The quantum walk procedure generated by our algorithm is shown in \autoref{fig:whole_circuit}. 

Recall the shift operator $S$ shown in \autoref{eq:S} in our quantum walk procedure, one can easily see that 
\[\begin{aligned}
    S&=\sum_{x,y=0}^{d-1}\left(\ket{x}\bra{x}\otimes\ket{0}\bra{0}+\ket{(x+1)\bmod{d}}\bra{x}\otimes\ket{1}\bra{1}\right)\\
    &\otimes\left(\ket{y}\bra{y}\otimes\ket{0}\bra{0}+\ket{(y+1)\bmod{d}}\bra{y}\otimes\ket{1}\bra{1}\right). 
\end{aligned}\]
This indicates that the shift operator is naturally a tensor product of two local operations respect to the two sides of our system. Thus, the only remaining operation to be dealt with is the coin operator $C^{(k)}$, which takes full control from the position space $\mathcal{H}_P$ to the coin space $\mathcal{H}_C$.

For the coin operators $C^{(k)}$ in our algorithm, recall that $C^{(k)}(x,y)$ is identity for all $0\le x,y<k$, and each $C^{(k)}$ is essentially controlled uniformly controlled gates (UCG), which is the generalization of controlled gate with different values of control qubits leading to different unitaries acting on target qubits. 

\begin{figure}[htbp]
    \centering
    \mbox{\Qcircuit @C=0.7em @R=1em{
        \lstick{p_1} &\rstick{x}\qw\qwx[1] &\qw & & &\rstick{k}\qw\qwx[1] &\rstick{k}\qw\qwx[1] &\rstick{x}\qw\qwx[1] &\qw\\
        \lstick{p_2} &\rstick{y}\qw\qwx[1] &\qw &\push{=} & &\rstick{y}\qw\qwx[1] &\rstick{k}\qw\qwx[1] &\rstick{k}\qw\qwx[1] &\qw\\
        \lstick{c_1} &\multigate{1}{C^{(k)}(x,y)} &\qw & & &\multigate{1}{C^{(k)}(k,y)} &\multigate{1}{C^{(k)}(k,k)} &\multigate{1}{C^{(k)}(x,k)} &\qw\\
        \lstick{c_2} &\ghost{C^{(k)}(x,y)} &\qw & & &\ghost{C^{(k)}(k,y)} &\ghost{C^{(k)}(k,k)} &\ghost{C^{(k)}(x,k)} &\qw
    }}
    \caption{$C^{(k)}$ can be treated as a UCG (left) and further decomposed into three UCG's (right).}
    \label{fig:coin_synthesis_1}
\end{figure}
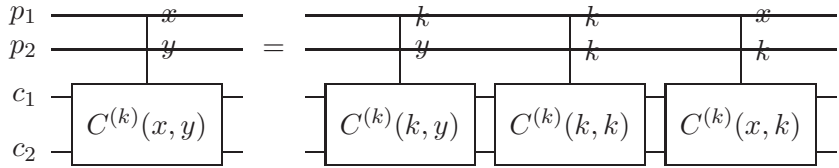

\begin{figure}[htbp]
    \centering
    \mbox{\Qcircuit @C=0.7em @R=1em{
        \lstick{p_1} &\rstick{k}\qw\qwx[2] &\qw &\qw &\qw &\rstick{x}\qw\qwx[3] &\qw &\rstick{k}\qw\qwx[2] &\qw\\
        \lstick{p_2} &\qw &\rstick{k}\qw\qwx[2] &\rstick{y}\qw\qwx[1] &\qw &\qw &\rstick{k}\qw\qwx[2] &\qw &\qw\\
        \lstick{a_1} &\targ &\qw &\ctrl{2} &\ctrl{1} &\qw &\qw &\targ &\qw\\
        \lstick{a_2} &\qw &\targ &\qw &\ctrl{1} &\ctrl{1} &\targ &\qw &\qw\\
        \lstick{c_1} &\qw &\qw &\multigate{1}{C^{(k)}(k,y)} &\multigate{1}{C^{(k)}(k,k)} &\multigate{1}{C^{(k)}(x,k)} &\qw &\qw &\qw\\
        \lstick{c_2} &\qw &\qw &\ghost{C^{(k)}(k,y)} &\ghost{C^{(k)}(k,k)} &\ghost{C^{(k)}(x,k)} &\qw &\qw &\qw
    }}
    \caption{Further transformation of the coin operator $C^{(k)}$. Two clean ancillary qubits $a_1$ (belonging to the first side of the system) and $a_2$ (belonging to the second side) are introduced. The initial states of the two ancillary qubits are both $\ket{0}$, and they end up with $\ket{0}$ too.}
    \label{fig:coin_synthesis_2}
\end{figure}
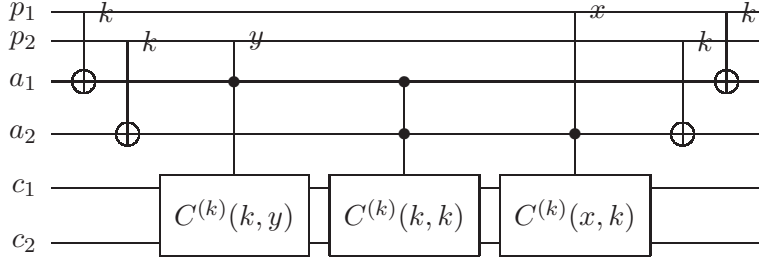

At step $k$, when the position space is in state $\ket{x,y}$, a unitary $C^{(k)}(x,y)$ is executed on the coin space and is shown in \autoref{fig:coin_synthesis_1}. The circuit representation of shift operator $S$ is similar.
After that, to achieve our goal, we introduce two qubit ancillae each belonging to one side of our distributed system. That is, expand the system to $\mathcal{H}_P\otimes\mathcal{H}_A\otimes\mathcal{H}_C\cong\left(\mathbb{C}^d\otimes\mathbb{C}^2\otimes\mathbb{C}^2\right)^{\otimes 2}$. With the help of these ancillary qubits, we can transform $C^{(k)}$ into equivalent circuits, as shown in \autoref{fig:coin_synthesis_2}.

\begin{figure}[htbp]
    \centering
    \mbox{\Qcircuit @C=0.7em @R=1em{
        \lstick{p_2} &\rstick{y}\qw\qwx[1] &\qw & & &\rstick{y}\qw\qwx[2] &\qw &\rstick{y}\qw\qwx[2] &\qw\\
        \lstick{a_1} &\ctrl{2} &\qw & & &\qw &\ctrl{1} &\qw &\qw\\
        \lstick{a_4} &\qw &\qw &\push{=} & &\targ &\ctrl{1} &\targ &\qw\\
        \lstick{c_1} &\multigate{1}{C^{(k)}(k,y)} &\qw & & &\qw &\multigate{1}{C^{(k)}(k,y)} &\qw &\qw\\
        \lstick{c_2} &\ghost{C^{(k)}(k,y)} &\qw & & &\qw &\ghost{C^{(k)}(k,y)} &\qw &\qw
    }}
    \caption{Decomposition of controlled the $C^{(k)}(k,y)$. We introduce another clean ancillary qubit $a_4$, belonging to the second side of the system. The initial and end state of $a_4$ are both $\ket{0}$.}
    \label{fig:coin_synthesis_3}
\end{figure}
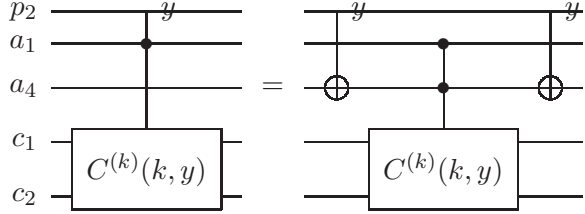

Finally, the only thing left is the decomposition of controlled $C^{(k)}(k,y)$, $C^{(k)}(k,k)$ and $C^{(k)}(x,k)$ gates. The controlled $C^{(k)}(k,k)$ gate is a $4$-qubit gate, which can be decomposed into single qubit gates and CNOT gates via standard techniques \cite{nielsen2002quantum}. For the controlled $C^{(k)}(k,y)$ gate, it's straightforward to see that one can decompose it into generalized Toffoli gates and $4$-qubit gates as shown in \autoref{fig:coin_synthesis_3}. The controlled $C^{(k)}(x,k)$ gate is similar. This results in $O(d^2)$ CNOT gates between the two sides $a_1,a_3,c_1$ and $a_2,a_4,c_2$ in total.

\subsection{Relationship with quantum state preparation}\label{ssec:QSP}
For quantum computation, both quantum walk and quantum circuit are universal computing models. Quantum state preparation is a well-studied problem under circuit model, which aims to generate a given state $\ket{\phi}=\sum_{x=0}^{2^n-1}\alpha_x\ket{x}$ from $\ket{0}^{\otimes n}$ by a quantum circuit. In general, this problem is of great essence and a series of works have been devoted to reducing the resources consuming such as circuit size, circuit depth and the number of ancillary qubits\cite{grover2002creating,bergholm2005quantum,plesch2011quantum,zhang2021low,sun2021asymptotically}. In particular, when one has $O(n)$ ancillary qubits, Sun et al give quantum circuits preparing any $n$-qubit states of $O(2^n)$ size and $O(\frac{2^n}{n})$ depth that reaches the lower bounds of size and depth simultaneously \cite{sun2021asymptotically}.

Now we show how our quantum engineering procedures via quantum walk can be used to design the quantum circuit for quantum state preparation with the same performance. Given a quantum state preparation instance of size $n$, we first divide the $n$ qubits into two parts of $\lceil0.5n\rceil$ each and treat the two parts as a bipartite system of dimension $2^{\lceil0.5n\rceil}$ (by adding an ancillary qubit if $n$ is odd). Then we utilize our scheme introduced in \autoref{sec:scheme} on this system. This results in a quantum circuit with mainly two kinds of operators: the coin operators $C^{(k)}$ and the shift operators $S$. We need to decompose them into single-qubit gates and CNOT gates.

The decomposition of coin operators is mainly based on a method decomposing a uniformly controlled gate (UCG) with $n$ controlled qubits and constant number of targets into quantum circuit with size $O(2^n)$ and depth $O(\frac{2^n}{n})$ \cite{sun2021asymptotically}. The coin operators $C^{(k)}$ can be seen as UCG's: as shown in \autoref{fig:whole_circuit}, when the state of the position space is $\ket{x,y}$, $C^{(k)}(x,y)$ is executed on the coin space. Utilizing the techniques mentioned above directly, each coin operator can be implemented by a quantum circuit of size $O(2^n)$, but this makes the whole circuit too big. In fact, we can do much better by using the equivalence shown in \autoref{fig:coin_synthesis_2}, in which the controlled $C^{(k)}(k,y),C^{(k)}(x,k)$ gates can be treated as one-qubit controlled UCG's. Given a quantum state preparation instance of size $n$, we first divide the $n$ qubits into two parts of $\lceil0.5n\rceil$ each and treat the two parts as a bipartite system of dimension $2^{\lceil0.5n\rceil}$ (by adding an ancillary qubit if $n$ is odd). Using the technique from \cite{sun2021asymptotically}, these UCG's can be decomposed into a quantum circuit of $O(2^{0.5n})$ CNOT and single-qubit gates which has $O(\frac{2^{0.5n}}{n})$ depth. As shown in \autoref{fig:coin_synthesis_2}, this results in a quantum circuit of $O(2^{0.5n})$ Toffoli and two-qubit gates which has $O(\frac{2^{0.5n}}{n})$ depth. Each Toffoli and two-qubit gate can be decomposed into CNOT and single-qubit gates of constant size and depth. Thus, the overall size and depth of the decomposition of $C^{(k)}(x,y)$ is $O(2^{0.5n})$ and $O(\frac{2^{0.5n}}{n})$ respectively.

The shift operator $S$ in this setting can be seen as controlled quantum adders. If one views the position space as two quantum registers, then different states of coin space control ``add one" operation on different registers. To be concrete, the first (second) register is added by one when the first (second) coin qubit is in state $\ket{1}$. Quantum adder is a well-studied object \cite{draper2000addition,cuccaro2004new,draper2004logarithmic}. In our application, each shift operator $S$ consists of two controlled ``add one" operation on $\lceil0.5n\rceil$ qubits, and this can be realized by a quantum circuit of size $O(n)$ and depth $O(\log n)$ \cite{draper2004logarithmic}.

To sum all these costs, according to \autoref{fig:whole_circuit}, there are $2^{\lceil0.5n\rceil}$ coin operators each of size $O(2^{0.5n})$ and depth $O(\frac{2^{0.5n}}{n})$, and $2^{\lceil0.5n\rceil}$ shift operators each of size $O(n)$ and depth $O(\log n)$. Thus, the overall size and depth of the circuit is $O(2^n)$ and $O(\frac{2^n}{n})$, which matches the result proposed in Ref. \cite{sun2021asymptotically} as promised.

\section{Discussion}\label{sec:discussion}
In this work, we put forward two schemes for engineering arbitrary multipartite high-dimensional quantum states via quantum walks with time- and position-dependent coins, and discuss several potential areas that our scheme can have applications in. In fact, it can be applied in other quantum information protocols where a particular multipartite high-dimensional entangled state is needed. Also, it raises theoretical support for potential physical experiments of quantum walk in multipartite systems since the both time- and position- dependent coin operation has been realized in real physical systems \cite{zhang2022arbitrary,hou2018deterministic}. On the other hand, we believe that this engineering task can also be accomplished by quantum walk with position-dependent coins which provide enough degree of freedom, thus it deserves further exploration. In general, we hope that this is a possible direction toward better understanding of the quantum information involved in multipartite entanglements.

\section*{Acknowledgments}
This work was supported in part by the National Natural Science Foundation of China Grants No. 61832003,  61872334, 61801459, and the Strategic Priority Research Program of Chinese Academy of Sciences Grant No. XDB28000000.

\bibliographystyle{alpha}
\bibliography{ref}

\newcommand{\etalchar}[1]{$^{#1}$}
\begin{thebibliography}{WLASR06}

\bibitem[AA03]{aaronson2003quantum}
Scott Aaronson and Andris Ambainis.
\newblock Quantum search of spatial regions.
\newblock In {\em 44th Annual IEEE Symposium on Foundations of Computer
  Science, 2003. Proceedings.}, pages 200--209. IEEE, 2003.

\bibitem[AFOV08]{amico2008entanglement}
Luigi Amico, Rosario Fazio, Andreas Osterloh, and Vlatko Vedral.
\newblock Entanglement in many-body systems.
\newblock {\em Reviews of Modern Physics}, 80(2):517, 2008.

\bibitem[AGJK20]{ambainis2020quadratic}
Andris Ambainis, Andr{\'a}s Gily{\'e}n, Stacey Jeffery, and Martins Kokainis.
\newblock Quadratic speedup for finding marked vertices by quantum walks.
\newblock In {\em Proceedings of the 52nd Annual ACM SIGACT Symposium on Theory
  of Computing}, pages 412--424, 2020.

\bibitem[Amb07]{ambainis2007quantum}
Andris Ambainis.
\newblock Quantum walk algorithm for element distinctness.
\newblock {\em SIAM Journal on Computing}, 37(1):210--239, 2007.

\bibitem[ASS20]{ahmad2020one}
Rashid Ahmad, Uzma Sajjad, and Muhammad Sajid.
\newblock One-dimensional quantum walks with a position-dependent coin.
\newblock {\em Communications in Theoretical Physics}, 72(6):065101, 2020.

\bibitem[BBC{\etalchar{+}}93]{bennett1993teleporting}
Charles~H Bennett, Gilles Brassard, Claude Cr{\'e}peau, Richard Jozsa, Asher
  Peres, and William~K Wootters.
\newblock Teleporting an unknown quantum state via dual classical and
  einstein-podolsky-rosen channels.
\newblock {\em Physical Review Letters}, 70(13):1895, 1993.

\bibitem[BFK09]{broadbent2009universal}
Anne Broadbent, Joseph Fitzsimons, and Elham Kashefi.
\newblock Universal blind quantum computation.
\newblock In {\em 2009 50th Annual IEEE Symposium on Foundations of Computer
  Science}, pages 517--526. IEEE, 2009.

\bibitem[BNP{\etalchar{+}}06]{banuls2006quantum}
M~C Banuls, C~Navarrete, A~P{\'e}rez, Eugenio Rold{\'a}n, and J~C Soriano.
\newblock Quantum walk with a time-dependent coin.
\newblock {\em Physical Review A}, 73(6):062304, 2006.

\bibitem[BVMS05]{bergholm2005quantum}
Ville Bergholm, Juha~J Vartiainen, Mikko M{\"o}tt{\"o}nen, and Martti~M
  Salomaa.
\newblock Quantum circuits with uniformly controlled one-qubit gates.
\newblock {\em Physical Review A}, 71(5):052330, 2005.

\bibitem[CDKM04]{cuccaro2004new}
Steven~A Cuccaro, Thomas~G Draper, Samuel~A Kutin, and David~Petrie Moulton.
\newblock A new quantum ripple-carry addition circuit.
\newblock {\em arXiv preprint quant-ph/0410184}, 2004.

\bibitem[CDQ{\etalchar{+}}18]{chen2018observation}
Chao Chen, Xing Ding, Jian Qin, Yu~He, Yi-Han Luo, Ming-Cheng Chen, Chang Liu,
  Xi-Lin Wang, Wei-Jun Zhang, Hao Li, et~al.
\newblock Observation of topologically protected edge states in a photonic
  two-dimensional quantum walk.
\newblock {\em Physical Review Letters}, 121(10):100502, 2018.

\bibitem[CGW13]{childs2013universal}
Andrew~M Childs, David Gosset, and Zak Webb.
\newblock Universal computation by multiparticle quantum walk.
\newblock {\em Science}, 339(6121):791--794, 2013.

\bibitem[Chi09]{childs2009universal}
Andrew~M Childs.
\newblock Universal computation by quantum walk.
\newblock {\em Physical Review Letters}, 102(18):180501, 2009.

\bibitem[CSL08]{chandrashekar2008optimizing}
C~Madaiah Chandrashekar, Radhakrishna Srikanth, and Raymond Laflamme.
\newblock Optimizing the discrete time quantum walk using a su (2) coin.
\newblock {\em Physical Review A}, 77(3):032326, 2008.

\bibitem[DKCK04]{durt2004security}
Thomas Durt, Dagomir Kaszlikowski, Jing-Ling Chen, and Leong~Chuan Kwek.
\newblock Security of quantum key distributions with entangled qudits.
\newblock {\em Physical Review A}, 69(3):032313, 2004.

\bibitem[DKRS04]{draper2004logarithmic}
Thomas~G Draper, Samuel~A Kutin, Eric~M Rains, and Krysta~M Svore.
\newblock A logarithmic-depth quantum carry-lookahead adder.
\newblock {\em arXiv preprint quant-ph/0406142}, 2004.

\bibitem[DLW{\etalchar{+}}21]{daiss2021quantum}
Severin Daiss, Stefan Langenfeld, Stephan Welte, Emanuele Distante, Philip
  Thomas, Lukas Hartung, Olivier Morin, and Gerhard Rempe.
\newblock A quantum-logic gate between distant quantum-network modules.
\newblock {\em Science}, 371(6529):614--617, 2021.

\bibitem[Dra00]{draper2000addition}
Thomas~G Draper.
\newblock Addition on a quantum computer.
\newblock {\em arXiv preprint quant-ph/0008033}, 2000.

\bibitem[GJC12]{gottesman2012longer}
Daniel Gottesman, Thomas Jennewein, and Sarah Croke.
\newblock Longer-baseline telescopes using quantum repeaters.
\newblock {\em Physical Review Letters}, 109(7):070503, 2012.

\bibitem[GPE{\etalchar{+}}19]{giordani2019experimental}
Taira Giordani, Emanuele Polino, Sabrina Emiliani, Alessia Suprano, Luca
  Innocenti, Helena Majury, Lorenzo Marrucci, Mauro Paternostro, Alessandro
  Ferraro, Nicol{\`o} Spagnolo, et~al.
\newblock Experimental engineering of arbitrary qudit states with discrete-time
  quantum walks.
\newblock {\em Physical Review Letters}, 122(2):020503, 2019.

\bibitem[GR02]{grover2002creating}
Lov Grover and Terry Rudolph.
\newblock Creating superpositions that correspond to efficiently integrable
  probability distributions.
\newblock {\em arXiv preprint quant-ph/0208112}, 2002.

\bibitem[GWZ{\etalchar{+}}21]{gong2021quantum}
Ming Gong, Shiyu Wang, Chen Zha, Ming-Cheng Chen, He-Liang Huang, Yulin Wu,
  Qingling Zhu, Youwei Zhao, Shaowei Li, Shaojun Guo, et~al.
\newblock Quantum walks on a programmable two-dimensional 62-qubit
  superconducting processor.
\newblock {\em Science}, 372(6545):948--952, 2021.

\bibitem[HTS{\etalchar{+}}18]{hou2018deterministic}
Zhibo Hou, Jun-Feng Tang, Jiangwei Shang, Huangjun Zhu, Jian Li, Yuan Yuan,
  Kang-Da Wu, Guo-Yong Xiang, Chuan-Feng Li, and Guang-Can Guo.
\newblock Deterministic realization of collective measurements via photonic
  quantum walks.
\newblock {\em Nature Communications}, 9(1):1--7, 2018.

\bibitem[IMG{\etalchar{+}}17]{innocenti2017quantum}
Luca Innocenti, Helena Majury, Taira Giordani, Nicol{\`o} Spagnolo, Fabio
  Sciarrino, Mauro Paternostro, and Alessandro Ferraro.
\newblock Quantum state engineering using one-dimensional discrete-time quantum
  walks.
\newblock {\em Physical Review A}, 96(6):062326, 2017.

\bibitem[Kad22]{kadiri2022steered}
Gururaj Kadiri.
\newblock Steered discrete-time quantum walks for engineering of quantum
  states.
\newblock {\em arXiv preprint arXiv:2205.04872}, 2022.

\bibitem[KBB02]{karimipour2002quantum}
Vahid Karimipour, Alireza Bahraminasab, and Saber Bagherinezhad.
\newblock Quantum key distribution for d-level systems with generalized bell
  states.
\newblock {\em Physical Review A}, 65(5):052331, 2002.

\bibitem[KFMS10]{keet2010quantum}
Adrian Keet, Ben Fortescue, Damian Markham, and Barry~C Sanders.
\newblock Quantum secret sharing with qudit graph states.
\newblock {\em Physical Review A}, 82(6):062315, 2010.

\bibitem[Kim08]{kimble2008quantum}
H~Jeff Kimble.
\newblock The quantum internet.
\newblock {\em Nature}, 453(7198):1023--1030, 2008.

\bibitem[KKB{\etalchar{+}}14]{komar2014quantum}
Peter Komar, Eric~M Kessler, Michael Bishof, Liang Jiang, Anders~S S{\o}rensen,
  Jun Ye, and Mikhail~D Lukin.
\newblock A quantum network of clocks.
\newblock {\em Nature Physics}, 10(8):582--587, 2014.

\bibitem[KW13]{kurzynski2013quantum}
Pawe{\l} Kurzy{\'n}ski and Antoni W{\'o}jcik.
\newblock Quantum walk as a generalized measuring device.
\newblock {\em Physical Review Letters}, 110(20):200404, 2013.

\bibitem[LCE{\etalchar{+}}10]{lovett2010universal}
Neil~B Lovett, Sally Cooper, Matthew Everitt, Matthew Trevers, and Viv Kendon.
\newblock Universal quantum computation using the discrete-time quantum walk.
\newblock {\em Physical Review A}, 81(4):042330, 2010.

\bibitem[LHA{\etalchar{+}}20]{lu2020quantum}
Hsuan-Hao Lu, Zixuan Hu, Mohammed~Saleh Alshaykh, Alexandria~Jeanine Moore,
  Yuchen Wang, Poolad Imany, Andrew~Marc Weiner, and Sabre Kais.
\newblock Quantum phase estimation with time-frequency qudits in a single
  photon.
\newblock {\em Advanced Quantum Technologies}, 3(2):1900074, 2020.

\bibitem[MBO{\etalchar{+}}18]{majury2018robust}
Helena Majury, Joelle Boutari, Elizabeth O’Sullivan, Alessandro Ferraro, and
  Mauro Paternostro.
\newblock Robust quantum state engineering through coherent localization in
  biased-coin quantum walks.
\newblock {\em EPJ Quantum Technology}, 5(1):1--12, 2018.

\bibitem[Mon16]{montero2016classical}
Miquel Montero.
\newblock Classical-like behavior in quantum walks with inhomogeneous,
  time-dependent coin operators.
\newblock {\em Physical Review A}, 93(6):062316, 2016.

\bibitem[Mon17]{montero2017quantum}
Miquel Montero.
\newblock Quantum and random walks as universal generators of probability
  distributions.
\newblock {\em Physical Review A}, 95(6):062326, 2017.

\bibitem[NC02]{nielsen2002quantum}
Michael~A Nielsen and Isaac Chuang.
\newblock Quantum computation and quantum information, 2002.

\bibitem[PB11]{plesch2011quantum}
Martin Plesch and {\v{C}}aslav Brukner.
\newblock Quantum-state preparation with universal gate decompositions.
\newblock {\em Physical Review A}, 83(3):032302, 2011.

\bibitem[PHB{\etalchar{+}}21]{pompili2021realization}
Matteo Pompili, Sophie~LN Hermans, Simon Baier, Hans~KC Beukers, Peter~C
  Humphreys, Raymond~N Schouten, Raymond~FL Vermeulen, Marijn~J Tiggelman,
  Laura dos Santos~Martins, Bas Dirkse, et~al.
\newblock Realization of a multinode quantum network of remote solid-state
  qubits.
\newblock {\em Science}, 372(6539):259--264, 2021.

\bibitem[SMS{\etalchar{+}}09]{schmitz2009quantum}
Hector Schmitz, Robert Matjeschk, Ch~Schneider, Jan Glueckert, Martin
  Enderlein, Thomas Huber, and Tobias Schaetz.
\newblock Quantum walk of a trapped ion in phase space.
\newblock {\em Physical Review Letters}, 103(9):090504, 2009.

\bibitem[SS10]{sheridan2010security}
Lana Sheridan and Valerio Scarani.
\newblock Security proof for quantum key distribution using qudit systems.
\newblock {\em Physical Review A}, 82(3):030301(R), 2010.

\bibitem[STY{\etalchar{+}}21]{sun2021asymptotically}
Xiaoming Sun, Guojing Tian, Shuai Yang, Pei Yuan, and Shengyu Zhang.
\newblock Asymptotically optimal circuit depth for quantum state preparation
  and general unitary synthesis.
\newblock {\em arXiv preprint arXiv:2108.06150}, 2021.

\bibitem[Suz16]{suzuki2016asymptotic}
Akito Suzuki.
\newblock Asymptotic velocity of a position-dependent quantum walk.
\newblock {\em Quantum Information Processing}, 15(1):103--119, 2016.

\bibitem[SZP{\etalchar{+}}21]{suprano2021dynamical}
Alessia Suprano, Danilo Zia, Emanuele Polino, Taira Giordani, Luca Innocenti,
  Alessandro Ferraro, Mauro Paternostro, Nicol{\`o} Spagnolo, and Fabio
  Sciarrino.
\newblock Dynamical learning of a photonics quantum-state engineering process.
\newblock {\em Advanced Photonics}, 3(6):066002, 2021.

\bibitem[TLF{\etalchar{+}}18]{tang2018experimental}
Hao Tang, Xiao-Feng Lin, Zhen Feng, Jing-Yuan Chen, Jun Gao, Ke~Sun, Chao-Yue
  Wang, Peng-Cheng Lai, Xiao-Yun Xu, Yao Wang, et~al.
\newblock Experimental two-dimensional quantum walk on a photonic chip.
\newblock {\em Science Advances}, 4(5):eaat3174, 2018.

\bibitem[UF10]{underwood2010universal}
Michael~S Underwood and David~L Feder.
\newblock Universal quantum computation by discontinuous quantum walk.
\newblock {\em Physical Review A}, 82(4):042304, 2010.

\bibitem[WEH18]{wehner2018quantum}
Stephanie Wehner, David Elkouss, and Ronald Hanson.
\newblock Quantum internet: A vision for the road ahead.
\newblock {\em Science}, 362(6412):eaam9288, 2018.

\bibitem[WHSK20]{wang2020qudits}
Yuchen Wang, Zixuan Hu, Barry~C Sanders, and Sabre Kais.
\newblock Qudits and high-dimensional quantum computing.
\newblock {\em Frontiers in Physics}, page 479, 2020.

\bibitem[WLASR06]{walborn2006quantum}
S~P Walborn, D~S Lemelle, M~P Almeida, and P~H Souto~Ribeiro.
\newblock Quantum key distribution with higher-order alphabets using spatially
  encoded qudits.
\newblock {\em Physical review letters}, 96(9):090501, 2006.

\bibitem[XSL09]{xue2009quantum}
Peng Xue, Barry~C Sanders, and Dietrich Leibfried.
\newblock Quantum walk on a line for a trapped ion.
\newblock {\em Physical Review Letters}, 103(18):183602, 2009.

\bibitem[XZQ{\etalchar{+}}15]{xue2015experimental}
Peng Xue, Rong Zhang, Hao Qin, Xiang Zhan, Z~H Bian, Jian Li, and Barry~C
  Sanders.
\newblock Experimental quantum-walk revival with a time-dependent coin.
\newblock {\em Physical Review Letters}, 114(14):140502, 2015.

\bibitem[YZG{\etalchar{+}}19]{yan2019strongly}
Zhiguang Yan, Yu-Ran Zhang, Ming Gong, Yulin Wu, Yarui Zheng, Shaowei Li, Can
  Wang, Futian Liang, Jin Lin, Yu~Xu, et~al.
\newblock Strongly correlated quantum walks with a 12-qubit superconducting
  processor.
\newblock {\em Science}, 364(6442):753--756, 2019.

\bibitem[ZYG{\etalchar{+}}22]{zhang2022arbitrary}
Rong Zhang, Ran Yang, Jian Guo, Chang-Wei Sun, Yi-Chen Liu, Heng Zhou, Ping Xu,
  Zhenda Xie, Yan-Xiao Gong, and Shi-Ning Zhu.
\newblock Arbitrary coherent distributions in a programmable quantum walk.
\newblock {\em Physical Review Research}, 4(2):023042, 2022.

\bibitem[ZYY21]{zhang2021low}
Xiao-Ming Zhang, Man-Hong Yung, and Xiao Yuan.
\newblock Low-depth quantum state preparation.
\newblock {\em Physical Review Research}, 3(4):043200, 2021.

\end{thebibliography}
\end{document}